\documentclass[11pt]{article}

\usepackage{amssymb}
\usepackage{graphicx}
\usepackage{caption}
\begin{document}

\title{A new concept for design of photonic integrated circuits with the ultimate density and low loss}

\author {J. Petrovic$^{1,2}$*, J. Krsic$^{2}$, J. J. P. Veerman$^{3}$, A. Maluckov$^{2}$\\
\emph{$^{1}$\small Institut f\"{u}r Optik und Atomare Physik, Technische Universit\"{a}t Berlin,}\\\emph{\small Straße des 17. Juni 135, 10623 Berlin, Germany}\\
\emph{$^{2}$\small "VINČA" Institute of Nuclear Sciences, University of Belgrade,}\\  \emph{\small 12-14 Mike Alasa, 11000 Belgrade, Serbia}\\
\emph{$^{3}$\small Fariborz Maseeh Department of Mathematics and Statistics,}\\
\emph{\small Portland State University, Portland, OR, USA}\\
\emph{\small *E-mail: jovanap@vin.bg.ac.rs}}

\date{}
\maketitle
\noindent\textbf{Abstract:} We challenge the current thinking and approach to the design of photonic integrated circuits (PICs) for applications in communications, quantum information and sensing. The standard PICs are based on directional couplers, that provide a wide range of functionalities but do not fully respond to the major technological challenges: massive parallelisation of transmission channels, low-energy dissipation and small footprint. We propose a new concept for design of PICs with the ultimate downscaling capability, the absence of geometric loss and a high-fidelity throughput. This is achieved by a periodic continuous-time quantum walk of photons through waveguide arrays that leverages on the simple and effective algebraic approach to engineering waveguide couplings. We demonstrate the potential of the new concept by reconsidering the design of basic building blocks of the information and sensing systems: interconnects, multiport  couplers, entanglement generators and interferometers. An extensive feasibility check in dielectric and semiconductor fabrication platforms confirmed this potential.\\

\noindent Photonic miniaturisation and integration have been a backbone of the modern communication, sensing and information processing systems ~\cite{ElshaariNatPhoton2020}. The utmost processing speed and transmission capacity, together with the miniature size within the coherence length of the readily available light sources, make the photonic integrated circuits (PIC) a competitive platform for the future quantum information processing.

\begin{figure}[ht!]
	\centering
	\includegraphics[width=\textwidth]{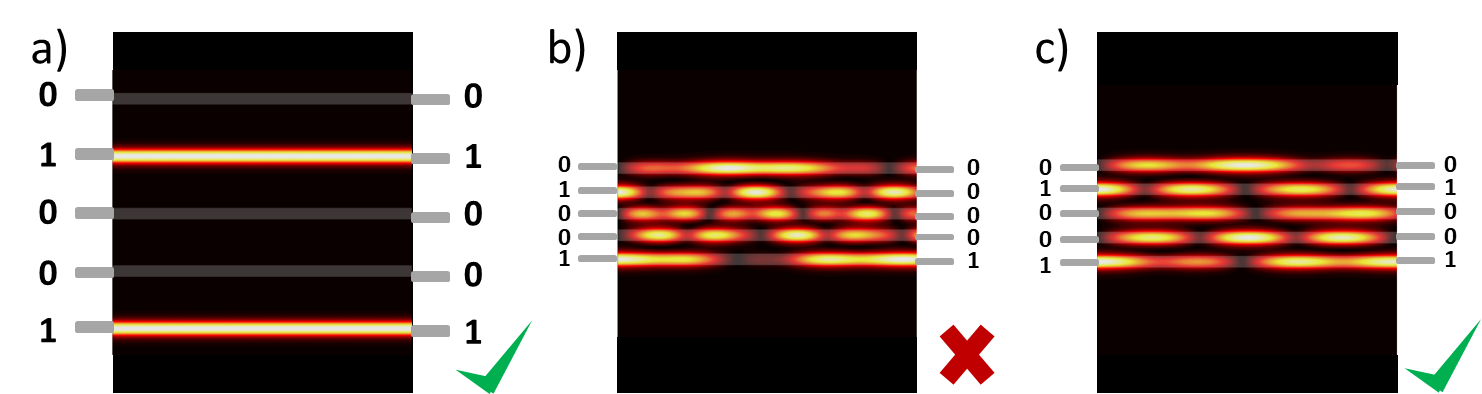}
\caption{\textbf{Crosstalk in interconnects.} a) A sparse uniform 5x5 port WGA with the negligible crosstalk transfers the light from left to right with high fidelity. b) Its densely packed variant experiences deleterious crosstalk, the input state is never repeated and the transfer fidelity is significantly reduced. c) A rearrangement of waveguides into a CWGA restores the high transfer fidelity while maintaining the small footprint. The fidelity is restored by the state revival at the output. The simulations were performed for the substrate refractive index $n=1.453$, index contrast $4.4\%$ and the waveguide diameter $3.5\mu m$, which ensures single mode guiding at 800 nm.}\label{fig:WGAvsCWGA}
\end{figure}

Multiplexing of waveguides on chips has been a workhorse and, at the same time, a foremost challenge before the competitive optical integration~\cite{SongNatComm2015}. The PICs stumble into the miniaturization wall due to the intrinsic crosstalk and the radiation loses at
waveguide bends~\cite{MillerNatPhoton2010}. The crosstalk is the unwanted coupling between waveguides~\cite{OkawachiPTL2011}. It affects the fidelity and bandwidth of the information transfer and becomes a forbidding factor in high-density optical interconnects~\cite{SongNatComm2015, HaurylauJSTQE06}, Fig.~\ref{fig:WGAvsCWGA}. Consequently, a variety of the crosstalk avoidance codes~\cite{QiaoJLT1994,XiongOX2020} and experimental techniques, such as the waveguide isolation~\cite{YangSciRep2020, BianSciRep2017, KwongPTL2014}, interlacing-recombination supercell~\cite{SongNatComm2015} and strong mode confinement by high refractive index contrast~\cite{UrinoOX2011, SmitSemicondSciTechnol2014}, have been developed.

\textit{Au contraire}, the possibility to control the coupling between waveguides allows for the engineering of the light propagation through waveguide arrays (WGAs), thus potentially bringing immense advantages in terms of parallelism and scalability in the number of modes. In the quantum picture, photons propagate through a WGA by performing the continuous-time quantum walk characterised by multiple path interferences. Since the seminal study by Childs~\cite{ChildsPRL2009} a lot of evidence has been provided that the quantum walk can be used as a tool for quantum computation. The viability of the WGAs as quantum devices has been demonstrated by the qubit interconnects~\cite{ChapmanNatComm2016}, entanglement generators~\cite{GraefeNatPhoton2014}, Boson sampling circuits~\cite{CarolanNatPhoton2014} and Bloch oscillators~\cite{PeschelOL1998, MorandottiPRL1999, KeilOptLett2012}. The corresponding WGA Hamiltonians have been constructed by inverting the evolution operator that describes a targeted functionality. However, the intricacy of the inverse design procedures has limited the solutions to the Hamiltonians with Wannier-Stark ladder spectra encountered in the condensed matter systems, namely, the $J_x$ spin rotations and Glauber-Fock lattices. However, to reach beyond the simulation of the known quantum systems, construction of new non-naturally-occurring Hamiltonians is required. A considerable effort is being put into solving the inverse problem by numerical optimisation~\cite{EndguldNJPQuantInf2018} and machine learning~\cite{YoussryQuantSciTechnol2020}. However, the particularity of these solutions and high requirements on the computational power arguably question the efficiency of the circuit design on classical computers. Hence, the absence of a streamlined design of the WGA-based PICs with the same underlying operating principle remains a major challenge before rivalling and surpassing the standard directional-coupler architectures in both functionality and scalability.

We respond to this challenge by introducing a new concept of PICs based on WGAs with commensurable eigenfrequencies (CWGAs). We show that it enables design of the PIC building blocks with the ultimate downsizing capability and the new functionalities relevant to the high-throughput communications, quantum state preparation and manipulation, and sensing. The pure algebraic design approach is corroborated numerically. Results of an extensive feasibility check in dielectric and semiconductor fabrication platforms are encouraging and are discussed briefly. Finally, we entertain the ideas of scalability in the number of dimensions and input photons, as well as the use of crosstalk and loss in multiparameter control.

\section*{Results}
\subsection*{Commensurable WGA}
The essential feature of a CWGA is that the ratio of any two of its eigenfrequencies is a rational number. It restricts the generally quasi-periodic light dynamics in WGAs to the highly-ordered periodic propagation~\cite{EfremidisOC05, CoutinhoLMA16, PetrovicAnnPhys2018}. In the tight-binding and the next-neighbour-coupling approximations, the fundamental mode in the $i^{th}$ waveguide can be modelled by a scalar complex wavefunction $\psi_i(z)$ and the CWGA by a tridiagonal model Hamiltonian $\mathbf{H}$. The Hamiltonian features the waveguide coupling coefficients on the side diagonals and the relative detunings between waveguide modes on the main diagonal. In the Schr\"{o}dinger picture, the light state $\psi(z)=(\psi_1(z),\ldots\psi_i(z),\ldots\psi_M(z))^T$, is evolved by the operator $\mathbf{T}=e^{-i\mathbf{H}z}$. The evolution operator describes the functionality of the array and can be related to its eigenspectrum by analytic expressions~\cite{PetrovicAnnPhys2018}. This enables construction of the Hamiltonians of the basic building blocks of PICs with freely chosen commensurate eigenfrequencies.

\subsection*{Interconnects}
High-fidelity optical interconnects are a premise for construction of compact classical and quantum optical chips and motherboards. The link capacity is maximized by the wavelength, time and spatial multiplexing. At high packaging densities, the crosstalk between channels (waveguides) distorts the spatially multiplexed signal, thus significantly reducing the transport fidelity and the link capacity~\cite{SongNatComm2015, HaurylauJSTQE06}. We propose the periodic transmission through CWGAs as the means of high-fidelity restoration of an input state at the output. This is achieved at the expense of discretising the viable output-port positions.

The state transfer through a CWGA interconnect is formally derived by selection of the identity evolution matrix, $T(kL)=I$, whereby the interconnect length equals $k$ multiples of the revival period $L$. A trivial algebraic analysis shows that all CWGAs fulfill the condition. For convenience and without loss of generality, we set $L=2\pi$ and represent Hamiltonian eigenspectrum by the integer eigenfrequencies $n_j,\,j=1,\ldots M$, thus arbitrarily sampling the Wannier-Stark eigenfrequency ladder. The free choice of eigenfrequencies marks a distinction from the commonly considered interconnects with equidistant eigenfrequencies~\cite{ChapmanNatComm2016, GordonOptLett2004, BellecOptLett2012}. The choice of an eigenspectrum dictates the ratios between the coupling coefficients but not their absolute values. This enables the CWGA construction by placing the identical waveguides at the designed interwaveguide separations, which can be scaled by an arbitrary number. The small scaling factors render high-density arrays.

\begin{figure}[ht!]
	\centering
\includegraphics[width=0.9\textwidth]{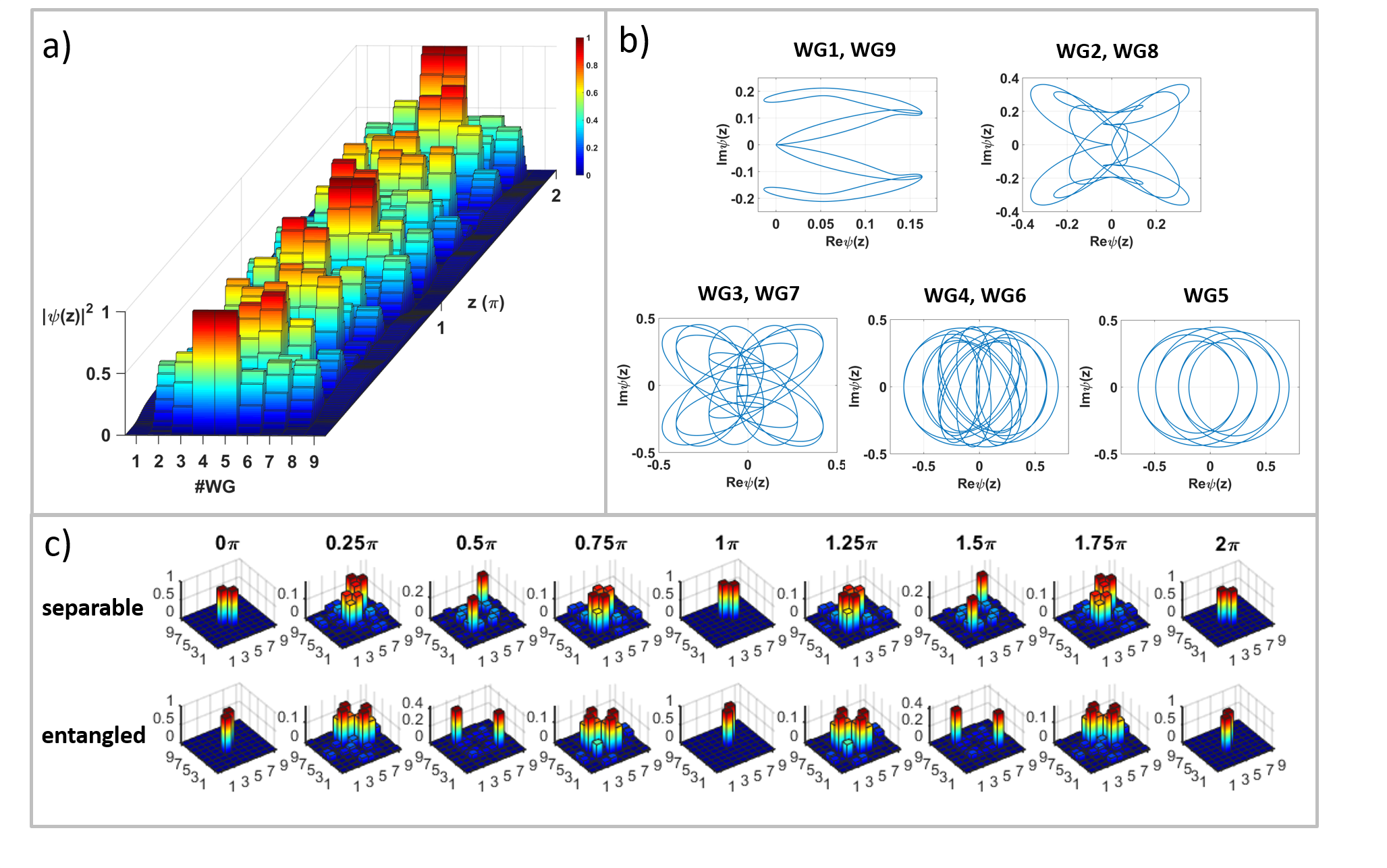}
\caption{\textbf{CWGA innterconnect.} Propagation of a photon pair through a 9-waveguide CWGA with eigenfrequencies $0,\,\pm 1,\, \pm 4,\, \pm 9,\, \pm14$. a) The photon number distribution along the array shows the revival at $z=2\pi$. b) The evolving mode wavefunctions $\psi_j,\,j=1,\ldots 9$ periodically sweep contours in the complex plane. This contrasts the mode wavefunctions in incommensurable WGAs, which evolve quasi-periodically and never repeat the same path. c) The revival of coincidence matrices demonstrates periodic restoration of photon correlations. The coincidence matrices also reveal the characteristic bunching of separable and antibunching of correlated photons.}\label{fig:Interconnects}
\end{figure}

The main signatures of the light propagation through a CWGA are the revivals of the photon number, relative phases between waveguide modes and photon correlations, Fig.~\ref{fig:Interconnects}. The photon-number revivals reach the maximum transfer fidelity $F=1$, whereby the fidelity is defined as \\$F=(\sum_{i=1}^M|\psi_{i}^{out}(z)||\psi_{i}^{target}|)^2$. The same result is obtained for the relative phases and the corresponding fidelity. Both types of revivals are evidenced by the periodic evolution of the mode wavefunctions $\psi_i(z),\,i=1,\ldots,M$ in the complex plane. The revivals of the photon correlations are observed in the evolution of the photon coincidences and can be further verified via a similarity measure between the coincidence matrices (Fig.~S2 in the Supplementary Information).

A CWGA with $M$ waveguides supports the transfer of $M/2$ dual-rail encoded qubits without the need for the customarily required perfect channel isolation~\cite{Burgarth2014}. Alternatively, a CWGA composed of $M\geq2^N$ or $M\geq d^D$ waveguides can act as an $M$ dimensional Hilbert space representable by $N$ qubits or $D$ qudits, encoded in $2$ and $d$ dimensional vectors, respectively. Finally, the absence of the bend-induced loss avails CWGAs to the single-rail encoding and hence to coupling to the stationary carriers of quantum information~\cite{DrahiQuantum2021}. Therefore, the commensurable interconnects can mitigate the downscaling bottleneck irrespectively of the encodings chosen for the future computing paradigm~\cite{WangFrontPhys2020}.

\subsubsection*{Perfect transfer}
Due to the fast development of the spin-based quantum computers, the perfect transfer through spin chains has attracted much attention~\cite{KayIJQI2010} and the WGAs have been used to simulate it \cite{BellecOptLett2012, ChapmanNatComm2016}. The CWGA interconnects represent a new family of the perfect transfer circuits, which do not rely on the nature-inspired coupled-system architectures. This opens up new possibilities for dealing with the transfer through the noisy or disordered channels.

The perfect transfer through a WGA with M waveguides is formally equivalent to the mirroring of the input state at the output, $\psi_{j}(z)=\psi_{M+1-j}(0)$, $j=\overline{1,M}$, and occurs at half-revival periods $z=(2k-1)\pi$, $k=1,2\ldots$. It is realized by the evolution operator equal to the exchange matrix, $T=J_M$, thus placing the restrictions on the parity of $M$ and the Hamiltonian eigenfrequencies $n_j$~\cite{EfremidisOC05}. In particular, the full transfer of a complex state is achieved for $M=4m+1$ where $m=1,2\ldots$, and the eigenfrequencies that alternate in parity when sorted in the ascending series initiated by an odd value. By allowing for the phase flip, as in the case of the perfect transfer through the spin chains, the suitable set of CWGAs is extended to those with $M=4m-1$ and the same eigenfrequency parity condition. For completeness, we note that the state mirroring in a CWGA with $M$ even requires degeneracy of its eigenfrequencies and hence the recursive decomposition of the CWGA into independent subarrays, which ends with a trivial solution $M=2$.

The results of the feasibility study described in Discussion indicate that the CWGA spectra and, hence, the coupling coefficients can be optimised to alleviate the effects of dephasing caused by the errors in waveguide separations and refractive indices. In particular, Fig.~\ref{fig:PTRobustness} therein shows that a single photon can be transferred through an optimised CWGA with a fidelity higher than that achievable in the equally disordered spin-chain, and that the solutions robust to disorder are not solitary.

\subsection*{Couplers}
Optical couplers combine the light inserted in the input ports to produce the targeted intensity redistribution at the output ports. The light propagating through the commonly used on-chip multi-mode interference (MMI) and path-interference directional couplers experiences output-port-insertion and bend-induced losses, respectively~\cite{BonneauNJP2012}. In the MMI it also experiences back reflections~\cite{SoldanoJLT1995}. The light propagating through the WGAs composed of single-mode waveguides experiences neither the losses nor back reflections. However, due to the challenging inverse design to match the MMI functionality, the WGAs have not been widely used as multiport couplers. The CWGAs radically change the situation as, owing to the highly ordered periodic light propagation, they render the conveniently round power ratios between the output ports.

\begin{figure}[ht!]
	\centering
\includegraphics[width=0.9\textwidth]{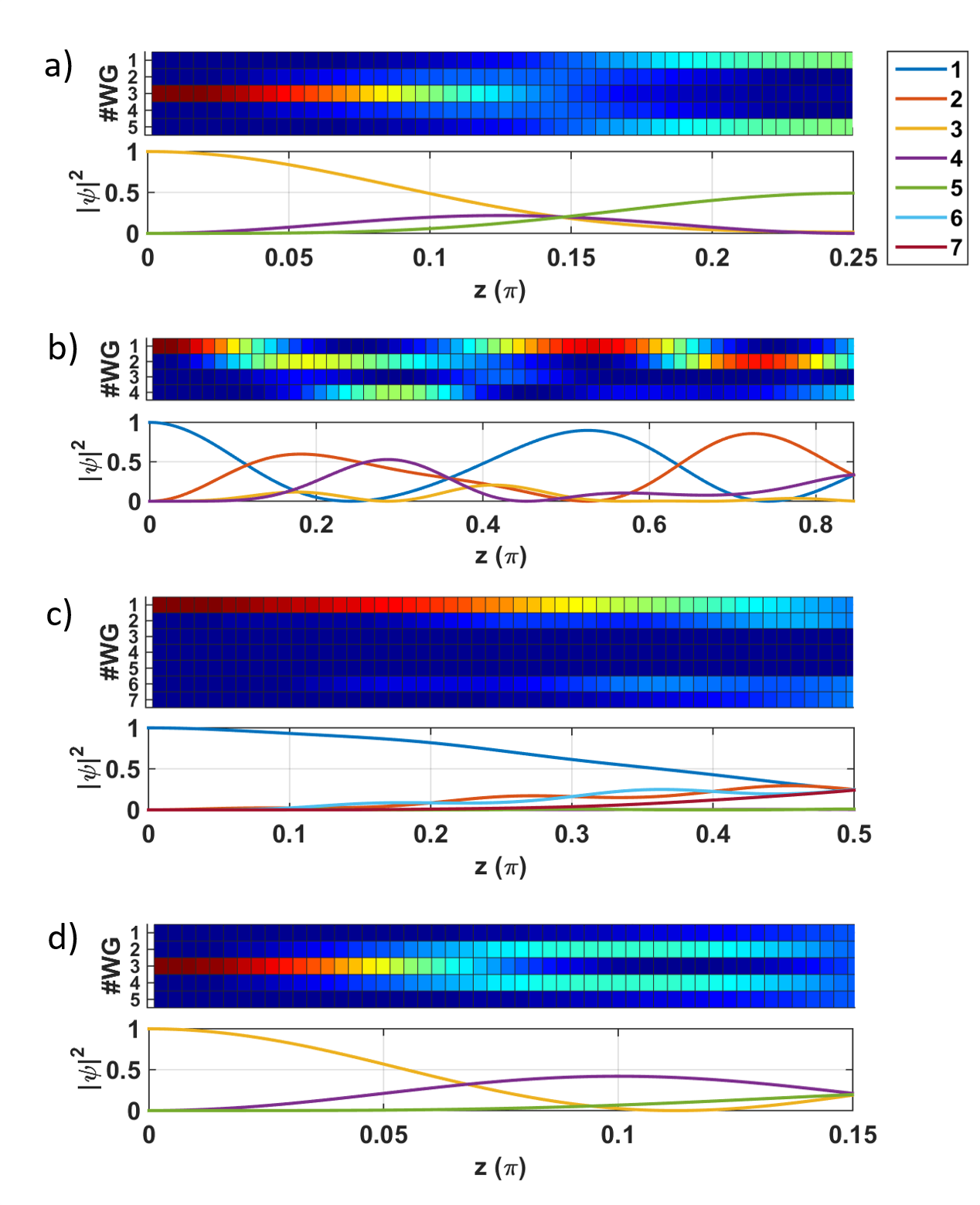}
\caption{\textbf{CWGA-based equal power dividers.} a) $1\times2$ power divider with $F=0.9844$ realised by a CWGA with eigenfrequencies $0,\,\pm 3,\,\pm 4$. b) $1\times3$ power divider with $F=0.9982$ realised by an asymmetric CWGA with eigenfrequencies $\pm 2,\,\pm 5$, and parameters $\varepsilon_1=1, \varepsilon_2=-1$ and $s=2.24$ \cite{PetrovicAnnPhys2018}. c) $1\times4$ power divider with $F=0.9752$ realised by a CWGA with eigenfrequencies $0,\,\pm 10,\,\pm 101$. d) $1\times5$ divider with $F=0.999999$ realised by a CWGA with eigenfrequencies $0,\,\pm 2,\,\pm 5$. $1\times5$ divider can also be realised by the CWGA in a), with $F=0.999968$. The same colour coding is applied in all figures and truncated to the number of waveguides as necessary.}\label{fig:Couplers}
\end{figure}

To demonstrate the versatility of the CWGA design, in Fig.~\ref{fig:Couplers}a)-d) we show a series of $1\times N$ equal-power dividers. The same divider can be realised in various ways, for example $1\times5$ splitting can be realised by the CWGAs in a) and d). On the other hand, a single CWGA can be used to produce dividers with different $N$ by targeted placing of the output ports. For example, 5-waveguide CWGA in a) can serve as $1\times5$ or $1\times2$ divider when the output is collected at $z=$0.146 $\pi$ or $z=$0.25 $\pi$, respectively. The choice of the input port is another option to select the power splitting ratio~\cite{PetrovicOptLett2015}. Particularly appealing is the design of asymmetric CWGAs, which is not fixed by the chosen system eigenspectrum. Rather, the asymmetric Hamiltonians feature one or more free parameters that can be used to tune the coupling ratios. This strategy was used to set the $1\times3$ splitting shown in b). Finally, by making use of the highly imbalanced coupling coefficients, the splitting ratios can be tuned to asymptotically approach the ideal values. Such couplers support robust boundary states, Fig.~\ref{fig:Couplers} c). The CWGA couplers are phase-sensitive and as such suitable for construction of bend-free interferometers (see the example of an unfolded Michelson interferometer in the Supplementary information). The planar geometry offers a new approach to the in-plane multi-arm interferometric sensing.

\subsection*{Entanglement generators}
\begin{figure}[ht!]
	\centering
\includegraphics[width=0.9\textwidth]{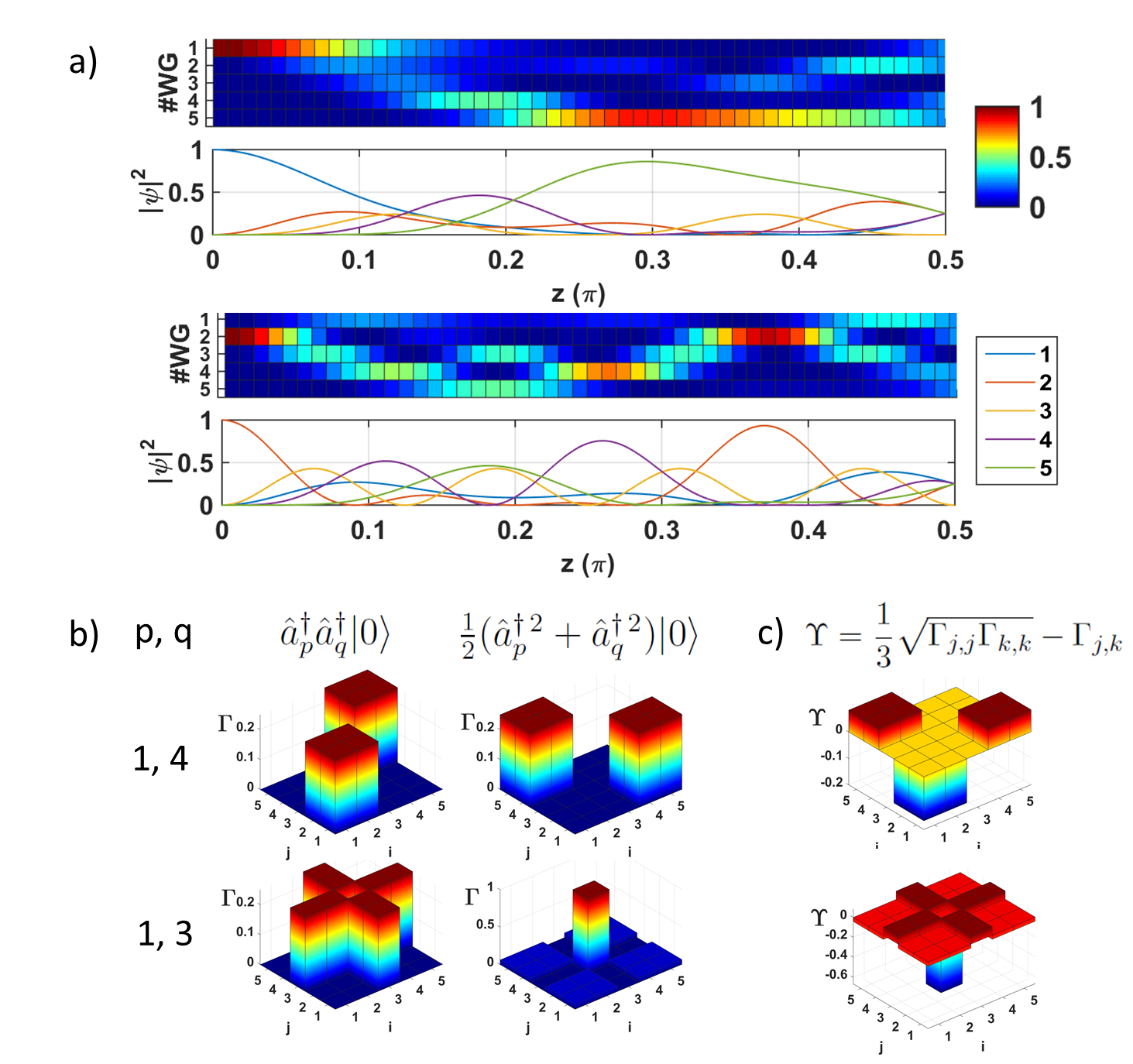}%
\caption{\textbf{W-state generator}. Shown is the CWGA with eigenfrequencies $0,\,\pm3,\,\pm8$ and the length equal to the quarter of the revival period. a) A single photon entering either of the ports 1, 2, 4 or 5 produces a $W_4$-state across the same four output ports.  b) Coincidence matrices for an indistinguishable photon pair inserted into ports $p,\,q$. For $p=1, q=4$, bunching and antibunching in the pairs of end waveguides are observed for the separable and correlated photons, respectively. The correlated photons entering ports $p=1,\,q=3$ (or $p=2,\,q=3$) bunch in the middle waveguide. c) Both cases of bunching shown in b) violate the Bell-like inequality $\Upsilon<0$.}\label{fig:W}
\end{figure}

Quantum interference of all possible photon paths within a WGA supports the generation of path-entanglement. The intrinsic parallelism of the WGAs is fully exploited in generation of multi-partite entangled W-states, that are known for high robustness to loss~\cite{DuerPRA2000}. The generalised W-state is given by a uniform distribution of a single photon over N waveguide modes and is conveniently represented in the Heisenberg picture by $|W_N\rangle=\frac{1}{\sqrt{N}}\sum^{N}_{k=1}\hat{a}_k^{\dag}|0\rangle
$, where $\hat{a}_k^{\dag}$ is the creation operator in waveguide $k$. The inverse WGA designs have been reported for generation of $W_N$ with an odd $N$,~\cite{Perez-LeijaPRA2013, PaulJO2014}. The $W$-states with an even $N$ have been designed by concatenation of directional couplers instead~\cite{GraefeNatPhoton2014}.

We show that the analytical design of CWGAs permits the generation of W-states with both even and odd $N$. Indeed, the Hamiltonians of the equal-power dividers $1\times N$ shown in Fig.~\ref{fig:Couplers} are also the solutions for $W_N$-state generators. Moreover, the mirror-symmetric 5-waveguide arrays with an odd eigenvalue $n_1$ and an even eigenvalue $n_2>n_1$ act as $W_4$-state generators when excited at port 2 or 4. If $n_2=4m,\,m=1,2\ldots$, the $W_4$-state can be generated also from a photon inserted into the port 1 or 5, Fig.~\ref{fig:W}. Besides the exact solutions, a number of asymptotic solutions can be easily derived in the form of CWGA with highly imbalanced coupling coefficients, as in Fig.~\ref{fig:Couplers}c). The $W_N$-state generators inherit from the couplers also an option to control the output state distribution by changing the input port and the position of the output ports. Here, these options can be used to choose the dimension $N$ of the output state. If necessary, the reduction in the array dimension is easily achieved by termination of the $M-N$ output ports that carry zero power.

CWGAs are a new playground also for the investigations of the effects of path entanglement on the multiphoton input states. To demonstrate this, we calculated the dynamics of correlations of two indistinguishable photons evolving through the CWGA in Fig.~\ref{fig:Interconnects} and the W-state generator in Fig.~\ref{fig:W}. The former supports the characteristic bunching and antibunching of the separable and correlated photons, respectively~\cite{GraefeJPB2020}. The latter, in addition, supports the photon clustering in the nearby modes at the end waveguides of the array. Interestingly, if a photon from a correlated pair is inserted into the middle waveguide of the W-state generator, it attracts its pair inserted elsewhere to bunch in the middle output port. In all cases of bunching, the non-classical correlations were confirmed by observing the violation of the Bell-like inequality~\cite{BrombergPRL2009}. Further scaling in the number of undistinguishable photons at the input leads to the parallelisation of quantum random walks, representable by synthetic lattices in the photon-number space~\cite{TschernigPR2019}.

Due to the commensurability, all quantum-interference induced effects repeat with a period shorter or equal to the revival period, Fig.~\ref{fig:Interconnects}.

\section*{Discussion}
A distinct advantage of the continuous walk in WGAs with respect to the discrete walk in directional couplers is the absence of radiation losses at waveguide bands. The losses in the typically used glasses set the lower bound of the bend radii to several millimeters (e.g.,15 mm at 800 nm), while the higher index contrast of semiconductor waveguides permits for an order of magnitude sharper bends (e.g., 600 $\rm\mu$m at 800 nm in silicon oxynitride)~\cite{PeruzzoScience2010}. The necessary bends and the associated branching leave significant chip areas redundant, thus curbing the scaling potential of the circuits~\cite{SmitAPLPhoton2019}. On the contrary, the commensurability enables WGA downscaling limited only by the conditions of light guiding. The commensurability concept withstands the breakdown of the tightly-binding and the nearest-neighbour approximations, whereby the inverse design calls for the numerical approach.

The obvious technological advantages of the low-loss systems have brought the Hermitian Hamiltonians into the focus of the work presented here. However, the recent research shows that the loss can be used as a valuable control parameter in the coupled waveguide systems whose real spectra are yielded by non-Hermitian PT-symmetric Hamiltonians~\cite{BenderPRL1998, JuarezPR2019}. Due to the absence of the bend-induced losses, CWGAs provide a possibility to independently control coupling and loss. The two-parameter control of CWGAs functionality merits further consideration and will be addressed elsewhere.

\begin{figure}[ht!]
	\centering
	\includegraphics[width=0.9\textwidth]{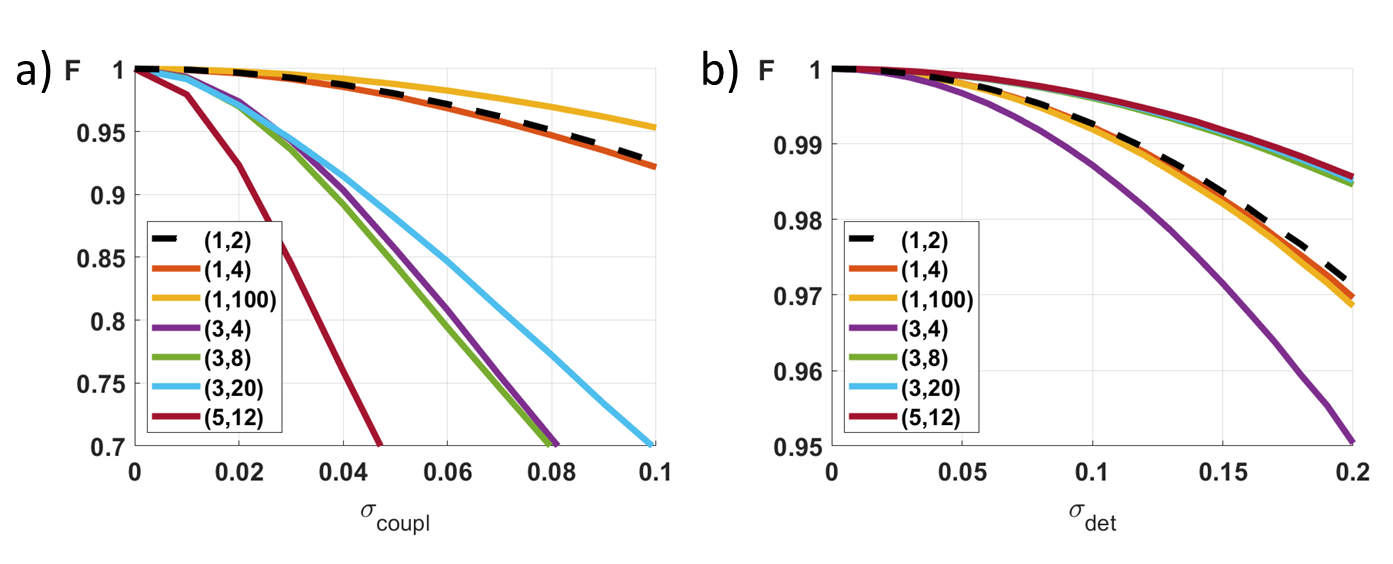}
\caption{\textbf{Robustness of the perfect transfer for different CWGA architectures.} The transfer fidelity decays with deviations in a) coupling coefficients, $\rm\sigma_{coupl}$, and b) phase detunings between waveguide modes, $\rm\sigma_{det}$. Shown is the mean fidelity obtained by averaging over $10^5$ points for each value of $\sigma$. It is assumed that all waveguides have the same normal distribution of parameters around their design values. The CWGA that emulates the spin coupling is shown by a dotted line to ease the comparison. The measure of disorder adopted here is general and pertains to all CWGA embodiments.}\label{fig:PTRobustness}
\end{figure}

PICs based on the CWGA building blocks have been envisaged as optical chips with integrated sources and detectors capable of addressing individual waveguides. The state-of-the-art CMOS technology~\cite{WangScience2018}, on-chip single-photon sources~\cite{SenellartNatNano2017} and superconducting nanowire single-photon detectors~\cite{PerniceNatComm2012} offer an optimistic view of the necessary technological developments.

Irrespectively of the chosen technological platform, the light propagation through CWGAs witnesses the system imperfections stemming from the fabrication tolerances in the waveguide refractive index profiles and separations. The former leads to the phase detuning representable by the diagonal disorder and the latter to the variations in coupling coefficients representable by the side-diagonal disorder. A comprehensive statistical study of disorder in the CWGA Hamiltonians has revealed that the consequent deviations from the periodic propagation pattern degrade the output quantum state fidelity. The degradation strongly depends on the CWGA architecture and the input conditions. Figure~\ref{fig:PTRobustness} compares fidelities of the end-to-end perfect transfer of a single photon through different CWGA architectures, including the spin-chain simulator~\cite{BellecOptLett2012}. All considered interconnects show remarkable robustness to the diagonal disorder and high sensitivity to the side-diagonal disorder. Optimisation of the array architecture yields a series of CWGAs, with the eigenfrequencies $n_1=1$ and $n_2$ even, that sustain the transfer fidelity above 0.999 in the presence of 1\% deviation of coupling coefficients or 20\% deviation of detuning. Therefore, while the spin-chain CWGA simulator has an excellent overall robustness, we are now in position to construct the new CWGA architectures that better respond to a particular source of dephasing. A further analysis was performed to determine the fabrication tolerance bounds in the presence of both disorders while taking into account the dependence on the input state. The `worst-case' study indicated that the transfer fidelity above 95\% can be sustained for the 0.3\% deviation of coupling coefficients and 10\% of detuning. The results unambiguously identify the waveguide positioning as the critical fabrication parameter. Applying the results to the CWGA shown in Fig.~\ref{fig:WGAvsCWGA}c) places the `worst case' scenario in the sub-10 nm precision regime for the 800 nm cut-off wavelength, at the border of the current semiconductor fabrication capabilities. On the other hand, the optimized CWGA architectures remain accessible to both semiconductor and dielectric processing technologies (see the Supplementary Information for more detail).

CWGAs are intrinsically scalable. Scaling up of the number of ports provides access to the high-dimensional Hilbert spaces suitable for operations on qudits, which have been marked as the next step towards the high-dimensional quantum computing~\cite{WangFrontPhys2020}. Further expansion of the commensurability concept to the second spatial dimension can be exploited in construction of the multi-layer on-chip interconnects as well as the spatially multiplexed fibre-to-chip interconnects~\cite{RadosavljevicJOSAB2015}. Finally, the passive CWGAs considered here lend themselves to reconfiguration and activation by standard techniques~\cite{BogaertsNature2020}, and hence to design of new monolithic all-WGA processors and flexible sensing platforms.

\section*{Methods}
\textbf{Inverse solutions.} The CWGA Hamiltonian composed of the waveguide coupling coefficients has been derived following the inverse procedure described in~\cite{PetrovicAnnPhys2018}. While the Hamiltonians of symmetric CWGAs are fully defined by the chosen eigenfrequencies, the Hamiltonians of asymmetric CWGAs can be tuned by analogue real parameters. The solutions discussed here are limited to those with the real coupling coefficients. The theory and graphs presented in Figs.~\ref{fig:Interconnects} -~\ref{fig:W} are based on the analytical model independent of the CWGA embodiment. On the other hand, numerical results in Fig.~\ref{fig:WGAvsCWGA} are obtained for a model WGA in a silicon oxynitride substrate~\cite{PeruzzoScience2010}. The strength of coupling coefficients was controlled by adjusting the separation between the identical waveguides as described in~\cite{BellecOptLett2012, PetrovicOptLett2015}. The wavefunction was evolved by the finite-difference beam-propagation method.

\noindent\textbf{Interconnects.} The transfer through interconnects is formally defined by the equivalence $\psi(z)=\psi(z=0)e^{i2k\pi}$, where $k=1,2,\ldots$ It is achieved by selection of the identity transfer matrix, $\mathbf{T}=I$, at each revival length $k2\pi$ along the array. A trivial analysis of the transfer matrix diagonalised in the form $\mathbf{T}^{diag}_{j,k}(2\pi)=e^{-in_j2\pi}\delta_{j,k}$, where $n_j$ are the eigenvalues of the Hamiltonian $\mathbf{H}$, shows that the equivalence condition is fulfilled only and only if all $n_j$ are integers.

\noindent\textit{Perfect transfer} can be mathematically described as mirroring of
the input state $\psi_{j}(z)=\psi_{M+1-j}(0),\,j=\overline{1,M}$ and realized by the exchange transfer matrix $\mathbf{J}_M$. Due to the required mirror symmetry of the transfer~\cite{KayIJQI2010}, it occurs at half-periods $z=(2k-1)\pi$, at which the CWGA transfer matrix diagonalizes in the form $\mathbf{T}^{diag}_{j,k}=(-1)^{n_j}\delta_{j,k}$ with $j,k=\overline{1,M}$. Here, the matrix equivalence condition $\mathbf{T}=\mathbf{J}_M$ is imposed in two steps. Firstly, the equivalence of the eigenspectra is employed to find the allowed number of waveguides in the array $M$. Secondly, the equivalence of the individual matrix elements is imposed to generate the necessary eigenvalue parity condition. We note that the similar results are obtained by the eigenvalue analysis applied to the fractional revivals in~\cite{EfremidisOC05}.\\
The condition of the spectral equivalence assumes different forms for different parities of $M$. For an odd $M$, the eigenspectrum of the exchange matrix $\mathbf{J}_M$ consists of $(M+1)/2$ eigenvalues $1$ and $(M-1)/2$ eigenvalues $-1$. Therefore, the WGA transfer matrix $T$ must possess, besides the compulsory eigenvalue $1$, an equal number $(M-1)/2$ of odd and even eigenvalues. Taking into account that the eigenspectrum is symmetric around zero, this imposes the condition on the total number of waveguides $M=4m+1,\,m=1,2,\ldots$. The element-by-element analysis of $T$ reveals a further necessary condition that the eigenvalues $n_j$ sorted in ascending order must be of alternating parity starting with an $odd$ $n_1$. If $M=4m-1$, it is possible to construct WGAs with $\mathbf{T}=-\mathbf{J}_M$ at half-periods and thus realise the transfer of amplitude with a phase flip of $\pi$.

\noindent For an even $M$, the eigenspectrum of the exchange matrix consists of $M/2$ pairs of eigenvalues $1, -1$. Its equivalence with the eigenspectrum of $T$ is possible only for the WGAs with an equal number of even and odd eigenvalues, yielding the condition $M=4m,\,m=1,2,\ldots$. However, a back-to-back comparison of $J_M$ and $T$ matrix elements shows that their equivalence requires existence of degenerate eigenvalues. Although formally viable, such commensurate WGAs have a number of coupling coefficients equal to zero and, hence, decompose to independent constitute arrays with smaller number of waveguides. The reduction recurs down to the trivial M=2 array.

\noindent\textbf{Coincidence counts.}
The coincidence counts at the output of a CWGA bear the signatures of photon correlations and are used routinely to characterise the path entanglement~\cite{GraefeJPB2020}. For a pair of distinguishable photons entering waveguides $p$ and $q$, the coincidence matrix has been calculated as $\Gamma_{j,k}=|T_{j,p}T_{k,q}|^2+|T_{j,q}T_{k,p}|^2$.
For indistinguishable separable photons $\hat{a}_p^{\dag}\hat{a}_q^{\dag}|0\rangle$, the coincidence matrix is $\Gamma_{j,k}=|T_{j,p}T_{k,q}+T_{j,q}T_{k,p}|^2$, and for the path-entangled photons
$\frac{1}{2}(\hat{a}_p^{\dag\,2}+e^{i\phi}\hat{a}_q^{\dag\,2})|0\rangle$,
$\Gamma_{j,k}=|T_{j,p}T_{k,p}+e^{-i\phi}T_{j,q}T_{k,q}|^2$, where $\hat{a}_p^{\dag}$ is a creation operator acting on waveguide $p$. The path entanglement violates the Bell-like inequality $\Upsilon=\frac{1}{3}\sqrt{\Gamma_{j,j}\Gamma_{k,k}}-\Gamma_{j,k}<0$.

\noindent\textbf{Disorder.} The disordered Hamiltonian was generated by assigning a random normal distribution with the relative standard deviation $\sigma_{\mathbf{H}_{i,j}}/\mathbf{H}_{i,j}=\sigma_{coupl},\,(j\neq i)$ to all side-diagonal elements and a random normal distribution with the relative standard deviation $\sigma_{\mathbf{H}_{i,i}}/\mathbf{H}_{i,i}=\sigma_{det}$ to all diagonal elements. Thereby, the coupling reciprocity $\mathbf{H}_{i,j}=\mathbf{H}_{j,i}$ was applied.

\section*{Acknowledgements}
J. Petrovic acknowledges support by the Berliner ChancengleichheitsProgramm (BCP). A. Maluckov and J. Krsic acknowledge  support  by  the  Ministry  of Education,  Science,  and  Technological  Development  of  the Republic  of  Serbia,  Grant  No.  451-03-68/2020-14/200017.  We thank U. Woggon, A. Szameit, A. P\'{e}rez-Leija and N. Stojanovic for fruitful discussions.



\newpage
\section*{SUPPLEMENTARY MATERIAL}
\section*{Interferometer}
\begin{figure}[htb]	\centering
\includegraphics[width=0.9\textwidth]{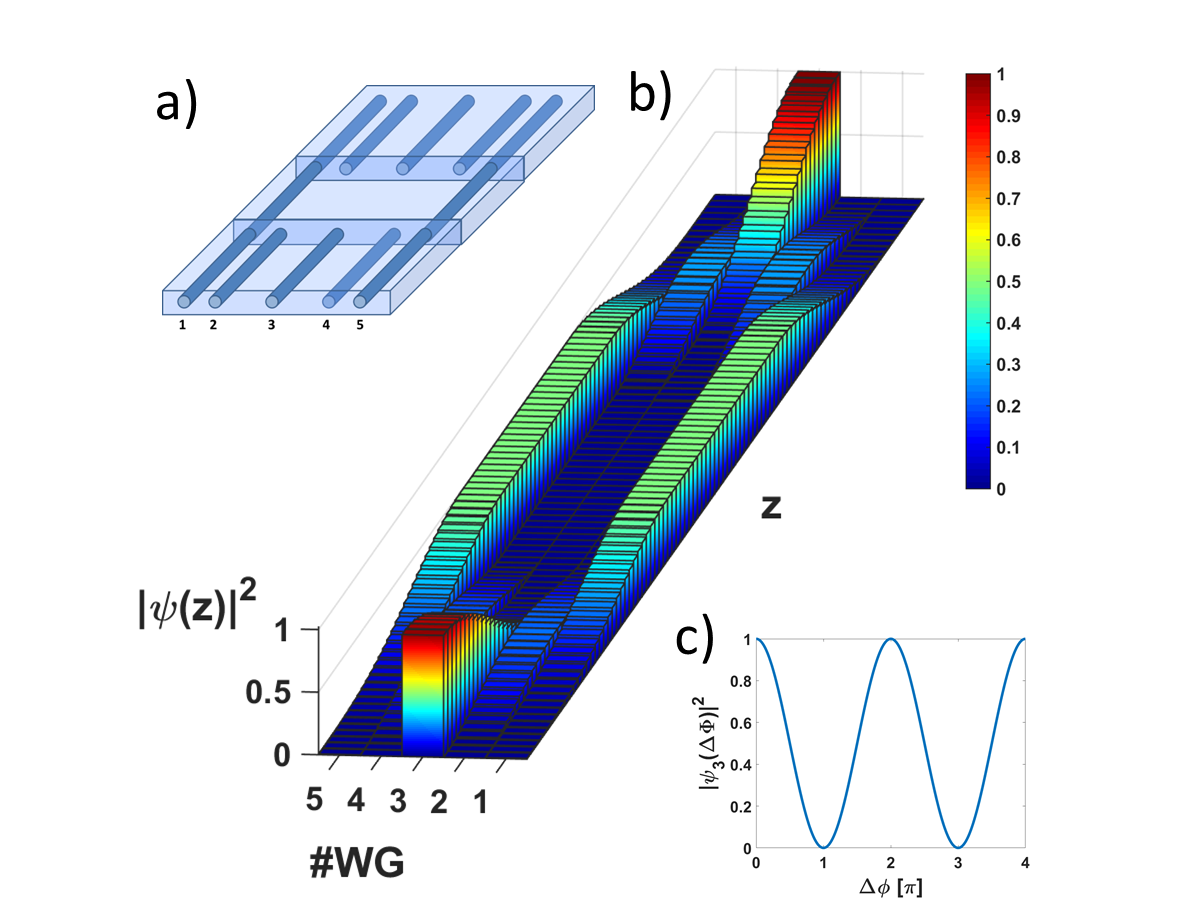}
\caption{a) \textbf{Michelson interferometer.} a) An interferometer composed of three CWGAs: two 5-waveguide arrays acting as couplers (with the eigenfrequencies $0,\,\pm2,\,\pm3$ and the length of $\pi/3$) and a 2-waveguide interconnect with negligible coupling between waveguides. b) Intensity of the coherent light propagating through the interferometer. c) The phase modulation in one arm of the interferometer generates the fringe pattern at the output.}\label{fig:Interferometer}

\end{figure}

The basic PIC building blocks presented in the article are used to construct an unfolded Michelson interferometer,  Fig.~\ref{fig:Interferometer}. It consists of two 1x2 couplers based on 5-waveguide arrays and a sandwiched interconnect based on a 2-waveguide array. Large separation of the WGs in the interconnect makes their coupling negligible, thus allowing for an independent phase change in each arm of the interferometer. The CWGA-based couplers enable construction of multiarm interferometers, known to enhance the measurement resolution \cite{PetrovicNJP2013}. Moreover, they support many-photon input states and entanglement across multiple modes, known to enhance the measurement sensitivity \cite{GiovannettiScience2004}.

\section*{Revival of photon correlations}
\begin{figure}[htb!]
	\centering
\includegraphics[width=\textwidth]{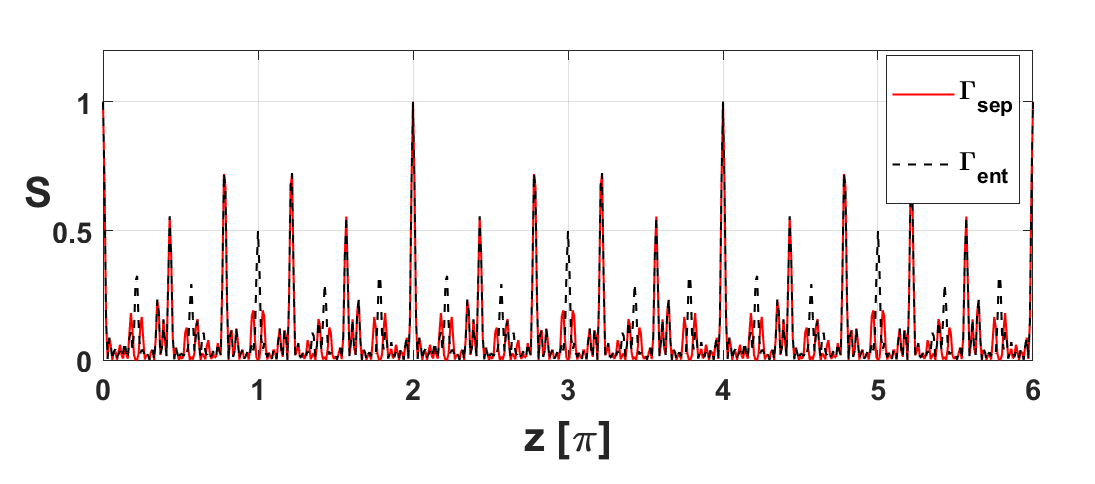}
\caption{\textbf{Photon-correlation revivals.} The similarity $S$ of the coincidence matrices at the input and along the CWGA from Fig.~2 for a pair of entangled ($\Gamma_{ent}$) and separable ($\Gamma_{sep}$) photons.}\label{fig:Similarity}
\end{figure}

The periodicity of photon correlations was assessed via the similarity measure $S$ that compares photon coincidence matrices along the array \cite{PeruzzoScience2010}. By definition, $S=\frac{(\sum_{j,k}\sqrt{\Gamma_{j,k}(0)\Gamma_{j,k}
(z)})^2}{\sum_{j,k}\Gamma_{j,k}(0)\sum_{j,k}\Gamma_{j,k}(z)}$, where the matrices $\Gamma_{j,k}$ are given in section Metods. Fig.~\ref{fig:Similarity} shows $S$ for the 9-waveguide array from Fig.~2. It confirms the full revivals of photon correlations at the multiples of revival periods $2\pi$, thus proving the periodicity of the path entanglement. The periodicity pertains to all CWGAs irrespective of the choice of the input ports and array architecture.

\section*{Dephasing}
\textbf{Disorder.} A comprehensive statistical study of the impact of the quenched (z-invariant) and unquenched disorder on the transfer fidelity was performed to estimate the CWGA fabrication tolerances. In the first step, the disorder is introduced in terms of the relative variations of the Hamiltonian matrix elements. This is a general model and pertains to any CWGA embodiment. In the second step, the dependence of the Hamiltonian matrix elements on the fabrication errors was incorporated.

\begin{figure}[ht!]
	\centering
\includegraphics[width=\textwidth]{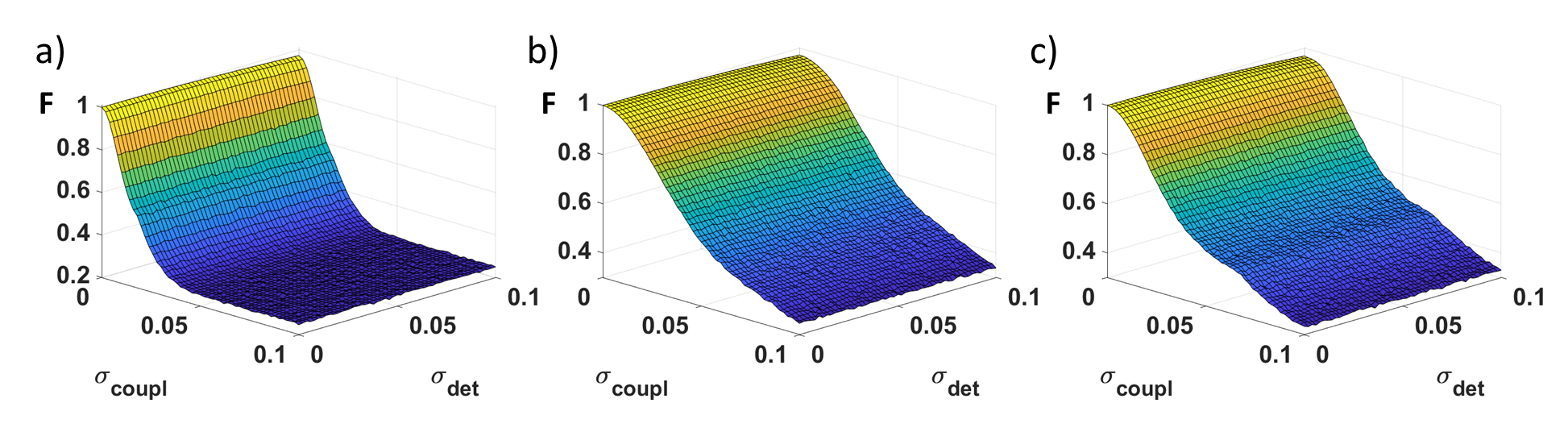}
\caption{\textbf{Disorder.} Impact of the diagonal ($\sigma_{det}$) and the side-diagonal ($\sigma_{coupl}$) disorder on the fidelity of the first full revival in CWGAs with the eigenfrequencies: a) $0, \pm 5, \pm 12$, b) $0, \pm 3, \pm 4$, and c) $0, \pm 1, \pm 4$}.
\label{fig:Dephasing}
\end{figure}

In the first step, the disordered Hamiltonian was generated by assigning the normal random distribution to all matrix elements around their design values. The relative standard deviation $\sigma_{\mathbf{H}_{i,j}}/\mathbf{H}_{i,j}=\sigma_{coupl},\,(j\neq i)$ was used for all side-diagonal elements and $\sigma_{\mathbf{H}_{i,i}}/\mathbf{H}_{i,i}=\sigma_{det}$ for all diagonal elements. Thereby, the coupling reciprocity $\mathbf{H}_{i,j}=\mathbf{H}_{j,i}$ was observed. The calculation was performed by the method developed for the quenched disorder in photonic networks \cite{thesis}. We found that the deterioration of the transfer fidelity depends on the input state and CWGA architecture and converges to dramatically different values in different cases. To ensure the `worst-case' scenario, we calculated the mean fidelity for every possible input state with single-rail encoding and reported the minima in Fig.~\ref{fig:Dephasing}. The representative CWGA architectures were chosen based upon the results in Fig. 5, namely, the CWGA with eigenfrequencies $0,\pm5,\pm12$ and poor robustness to the side-diagonal disorder and the CWGA with eigenfrequencies $0,\pm3,\pm4$ and poor robustness to the diagonal disorder. The control case was the CWGA with the eigenfrequencies $0,\pm1,\pm4$ and solid robustness to both disorders. In terms of the CWGA functionality, the two cases were considered: the perfect transfer to $z=\pi$ and the full revival at $z=2\pi$. The silicone oxynitride platform from Fig. 1 was used as the CWGA embodiment.

In the second step, the relative standard deviations of the side-diagonal and diagonal elements were translated to the deviations of waveguide separations and the refractive index contrast, respectively. The relative deviations of the coupling coefficients were calculated using the exponential dependence $\mathbf{H}_{i,j}=A e^{-\alpha d_{i,j}}$ on the waveguide separations $d_{i,j}$. The parameters $A$ and $\alpha$ depend on the waveguide embodiment. The diagonal disorder is represented by small differences in the mode propagation vectors caused by the errors in the waveguide refractive index profile. We approximated them by the relative errors in the refractive index contrast.

Results show that the quenched and non-quenched disorder have the qualitatively and quantitatively similar impact. Therefore, the fidelity reduction for the full revival is twice larger than for the perfect transfer, that takes one half of the revival period to develop. It what follows, we discuss the impact of the quenched disorder on the full revival.

The corresponding results in Fig.~\ref{fig:Dephasing} show that the side-diagonal disorder reduces the transfer fidelity by two orders of magnitude faster than the diagonal disorder. In particular, to reach the fidelity $F>0.95$, the coupling-coefficient tolerance should not exceed 0.35\%, while the detuning margin remains above 10\%. The dominance of the side-diagonal disorder, further amplified by the exponential dependence of the coupling coefficient on the waveguide separation, unambiguously defines the waveguide separation as the critical fabrication parameter. In the worst-case scenario, the corresponding upper bound of the waveguide-separation tolerance is 9 nm and of the refractive-index-contrast tolerance 0.1. On the other hand, the optimised CWGAs tolerate the errors of several tens of nanometers in waveguide separation. The tolerance increases further with decreasing index contrast.

The results indicate that the optimised CWGA architectures can be accessed by the cost-effective table-top techniques. A very successful table-top technique with respect to this is the laser writing, which has been widely used for fabrication of the CWGA spin-chain simulators \cite{BellecOptLett2012, SzameitJPB2008, GraefeJPB2020}, as well as directional-coupler based PIC \cite{SpringScience2013,CrespiNatPhoton2013, TillmannNatPhoton2013}. On the other hand, fabrication of the CWGA architectures with higher sensitivity to disorder, which also coincides with the higher index contrast, requires access to the semiconductor fabrication plants.\\
\begin{figure}[ht!]
	\centering
\includegraphics[width=0.9\textwidth]{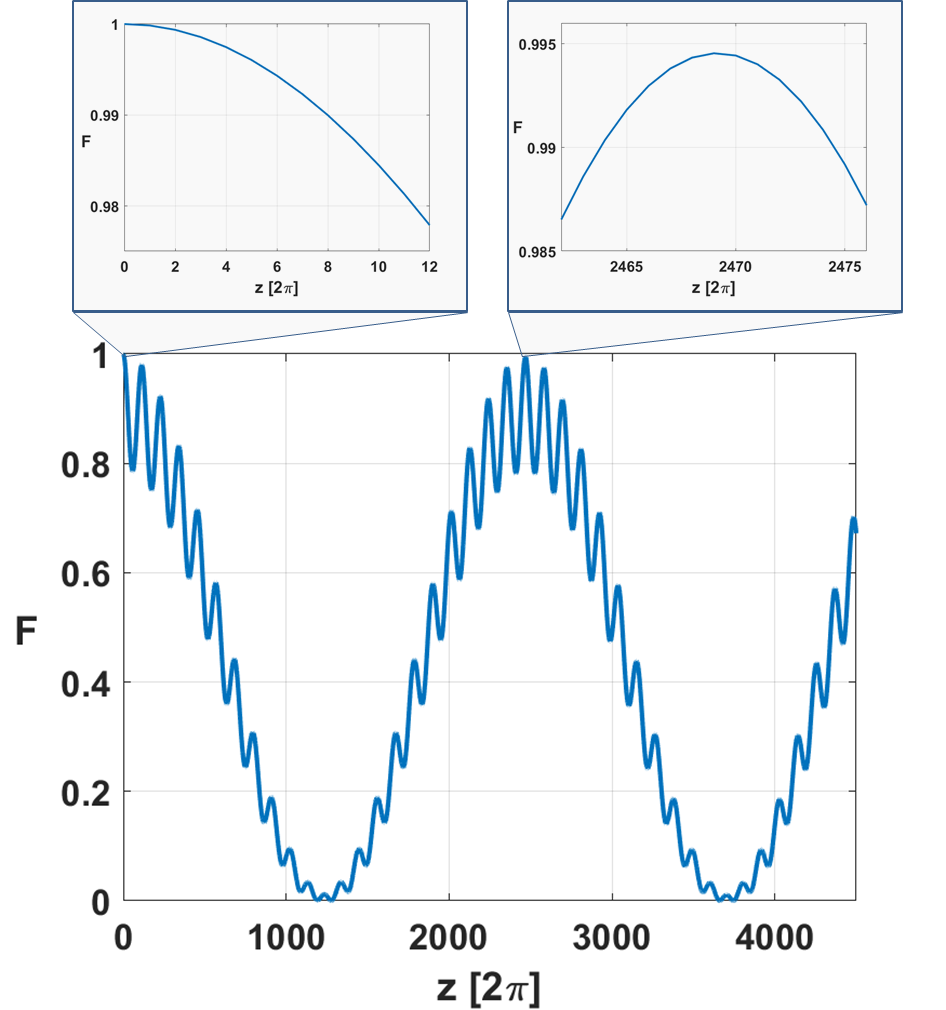}
\caption{a) \textbf{The long-range coupling}. Fidelity of the revivals of a single photon propagating through the CWGA from Fig.~1c) with the nearest-neighbour and the next-to-nearest-neighbour coupling. The photon is inserted into the input port 1 or 5. Only the full revivals at $z=2\pi k, \,k=1,2,\ldots$ are shown. The upper graphs zoom in on the fidelity maxima, evidencing its slow decay and the incomplete revivals.}
\label{fig:NNN}
\end{figure}

\noindent\textbf{Long-range coupling.} While the disorder can be curbed by improving the fabrication precision, the long-range evanescent field coupling to the non-nearest neighbours is inherent to the system. It does not invalidate the commensurability concept proposed but sets the limit to the validity of the tridiagonal model Hamiltonian. To appreciate the extent of this limitation, we model the next-to-nearest-neighbour coupling in the CWGA from Fig.~1c). The additional coupling was modelled by the further two side diagonals in the Hamiltonian. The corresponding coupling coefficients were calculated as $\mathbf{H}_{13}=\mathbf{H}_{35}=\mathbf{H}_{12}\mathbf{H}_{23}/A$ and $\mathbf{H}_{24}=\mathbf{H}_{23}^2/A$, and hence depend on the CWGA embodiment \cite{PetrovicOptLett2015}. As the coupling strength exponentially decays from the nearest towards further neighbours, the long-range coupling acts as a perturbation on the eigenfrequencies of the tridiagonal Hamiltonian and generates a slow quasi-periodic modulation of the light propagation pattern. Its impact on fidelity was calculated at each revival period over thousands of periods. The result in Fig.~\ref{fig:NNN} shows that it takes more than 8 revival periods for the fidelity to drop for 1\% and around 48 revival periods to drop to 80\%. The light dynamics in the first revival period, the most relevant for the operations on chip, remain practically intact. This is further evidenced by the numerical simulation in Fig.~1c) which implicitly includes all long-range couplings.

\end{document}